Revision 1

# STEM investigation of exsolution lamellae and "*c*"-reflections in Ca-rich dolomite from the Platteville Formation, West Wisconsin

Zhizhang Shen, Hiromi Konishi, Philip E. Brown, and Huifang Xu[*]

NASA Astrobiology Institute, Department of Geoscience,

University of Wisconsin - Madison

Madison, Wisconsin 53706

* Corresponding author: Dr. Huifang Xu

Department of Geoscience,

University of Wisconsin-Madison

1215 West Dayton Street, A352 Weeks Hall

Madison, Wisconsin 53706

Tel: 1-608-265-5887

Fax: 1-608-262-0693

Email: hfxu@geology.wisc.edu

## ABSTRACT

Dolomite crystals in partially dolomitized limestone from the Platteville Formation are both compositionally and microstructurally heterogeneous. A single dolomite crystal usually contains three phases: the host Ca-rich dolomite ($Ca_{1.14}Mg_{0.86}(CO_3)_2$), an Fe-bearing dolomite ($Ca_{1.06}Mg_{0.80}Fe_{0.14}(CO_3)_2$) and calcite inclusions. These three phases show similar orientations. The Ca-rich dolomite exhibits modulated microstructures with wavelength ranging from 7 to 30 nm. The modulated microstructures are not evident in Fe-bearing dolomite.

Modulations in the Ca-rich dolomite have three predominant orientation ranges in the studied sample: from (205) to (104), from (001) to ($\bar{1}01$), and ~ (110), which are consistent with previous studies. Bright-field (BF) and high angle annular dark-field (HAADF) images confirm that these modulations are due to chemical variation rather than strain or diffraction contrast. The Ca-rich lamellae are Mg-rich calcite with compositions ranging from $Ca_{0.85}Mg_{0.15}CO_3$ to $Ca_{0.70}Mg_{0.30}CO_3$. The observed results indicate that these Ca-rich exsolution lamellae formed during diagenesis. In this study, three kinds of "$c$"-reflections, which are weak spots in the halfway position between the principal reflections along the (104)*, ($\bar{1}12$)* and (110)* directions, have been found in the diffraction patterns of some Ca-rich dolomite. Mg-Ca ordering in $x$-$y$ planes was not observed directly in Z-contrast images. FFT patterns from the Z-contrast images do not show "$c$"-reflections. STEM images confirm that the "$c$"-reflections could result from multiple diffraction between the host dolomite and twinned Mg-calcite nano-lamellae under TEM imaging and diffraction modes.

# INTRODUCTION

The structure of a dolomite crystal ($R\bar{3}$) is similar to that of a calcite ($R\bar{3}c$) but with Ca and Mg layers alternating along the *c*-axis. The large differences in size between the $Ca^{2+}$ and $Mg^{2+}$ ions (33%) causes the cation ordering along the *c*-axis. With the nonequivalence of Ca and Mg layers, the symmetry is reduced from $R\bar{3}c$ to $R\bar{3}$. Many natural dolomites have an excess of $Ca^{2+}$, with composition up to ~ $Ca_{1.2}Mg_{0.8}(CO_3)_2$ which is quite different from the stoichiometric dolomite $CaMg(CO_3)_2$ (Reeder,1983; Reeder, 1992; Warren, 2000). The extra Ca in the dolomite structure causes an increase in the unit cell parameter and hence in *d*-spacings, since the radius of $Ca^{2+}$ is larger than that of $Mg^{2+}$ (The ionic radii for $Ca^{2+}$ and $Mg^{2+}$ in 6-fold coordinated are 1.00Å and 0.72Å respectively (Shannon,1976)). In Ca-rich dolomite, heterogeneous microstructures such as modulations and ordered superstructures have been reported (Reeder, 1992).

Finely modulated microstructure in ancient calcian dolomite was first noticed by Reeder et al. in 1978, and is very common in calcian dolomite, some calcite and calcian ankerite (Reeder and Wenk, 1979; Reeder, 1981; Gunderson and Wenk, 1981; Van Tendeloo et al., 1985; Reeder and Prosky, 1986; Miser et al., 1987; Reksten, 1990a; Wenk et al., 1991; and Fouke and Reeder, 1992). Modulation can either be pervasive throughout a crystal, or intergrown with areas devoid of modulation. This modulation was ascribed to be compositional fluctuation associated with excess Ca in dolomite by Reeder (1981), who initially interpreted it to have arisen by reorganization in the solid state. However, some modulations were believed to have formed during growth, based on the fact that the orientations of modulations are different in different growth sectors of a dolomite crystal (Reeder and Prosky, 1986; Miser et al., 1987; Fouke and Reeder, 1992).

Four types of reflections have been found in the diffraction patterns of dolomite. "*a*" reflections refer to reflections that existed in the diffraction pattern of calcite. The reflections which are found in dolomite but absent in calcite are termed "*b*" reflections. "*c*" reflections are very weak and usually streaked spots halfway between the principle reflections along any of the three directions of (110)*, (104)* and (012)* in the diffraction patterns of some dolomites (Reeder, 1981). The "*c*" type reflections are usually associated with the modulated microstructures in Ca-rich dolomite (Reeder and Wenk, 1979; Reeder, 1981; Van Tendeloo et al., 1985; Wenk and Zhang, 1985; Reksten, 1990a; Wenk et al., 1991; and Fouke and Reeder, 1992; Schubel et al., 2000). The "*d*" reflections are satellites around "*a*" and "*b*" reflections, which were found in a few dolomite samples (Wenk and Zenger, 1983).

Although microstructures in dolomite have been studied for the past few decades, there are still debates over the causes of modulated microstructures and "*c*" reflections. With the development of the technology of the electron microscope, a spherical aberration-corrected scanning transmission electron microscope (STEM) can image single atoms directly with sub-Å (less than 0.1nm) spatial resolution and gather chemical and structural information using high-angle scattered non-coherent electrons (Kirkland, 1998). The purpose of this study is to investigate the microstructures in Ca-rich dolomite from the Ordovician Platteville Formation in western Wisconsin using this technique.

**SAMPLES**

The rock sample collected from a partially dolomitized limestone outcrop in Prairie du Chien, Wisconsin, belongs to the Platteville Formation of the middle Ordovician Sinnipee Group. Euhedral dolomite crystals are found to grow in the micritic calcite matrix. The bulk rock sample was characterized by X-ray diffraction. Phase quantification using the Rietveld method is implemented in the Jade 9.0 Whole Pattern Fitting (WPF) program. The Rietveld calculations show that the dolomitized limestone contains 46% calcite, 42% dolomite, 5% quartz and 7% alkali feldspar by weight. In backscattered electron (BSE) images, euhedral dolomite crystals were composed of several compositionally distinct domains: BSE-bright unreplaced residual calcite, gray Fe-bearing dolomite grains and the dark host dolomite crystal. Specimens for STEM measurements were selected from areas containing large euhedral Ca-rich dolomite crystals extracted from doubly polished thin sections and then ion milled.

**EXPERIMENTAL METHODS**

The average chemical compositions of the host dolomite and Fe-bearing dolomite determined by EMPA are $Ca_{1.14}Mg_{0.86}(CO_3)_2$ and $Ca_{1.06}Mg_{0.80}Fe_{0.14}(CO_3)_2$ respectively. Ca, Mg, Fe and Mn were measured with a CAMECA SX51 instrument using wavelength-dispersive spectrometers at an accelerating voltage of 15 kV, a 10 nA beam current, and an interaction volume of ~3 μm. Calcite, dolomite, siderite and rhodochrosite were used as standards for Ca, Mg, Fe and Mn respectively.

The microstructures and interface structure between the inclusions and the host dolomite crystal were examined by using a spherical aberration-corrected FEG- STEM (Titan 80-200) operating at 200 kV at the University of Wisconsin-Madison. This instrument can image single atoms with ~ 0.08 nm spatial resolution in STEM mode. Probe current was set at 24.5 pA. Collection angle of HAADF detector for acquiring all the Z-contrast images ranges from 54 to 270 mrad (Corresponding to 7.5 (1/Å) to 38.2 (1/ Å) in reciprocal space).

Geometrically, the STEM can be regarded as an inverted Conventional TEM (CTEM). In STEM, different detectors sample different parts of the scattering space. The bright field (BF) detector usually collects over a small disc of low-angle coherently scattered coherent electrons centered on the optic axis of the microscope, whereas the high-angle annular dark field (HAADF) detector collects over an annulus of high-angle incoherently scattered electrons (Kirkland, 1998; Nellist, 2007).

The intensity in a HAADF image is strongly related to atomic number (Z) through the $Z^{1.7}$ dependence of the Rutherford scattering cross-section. An atomic resolution HAADF image is also called a Z-contrast image. The Z-contrast imaging technique can avoid multiple diffractions that commonly occur in HRTEM and electron diffraction modes that use elastic coherent electrons. The BF image is expected to contain diffraction and strain contrast that is less evident in the HAADF image.

**RESULTS**

Lamellae of bright contrast parallel to ~ (110) with wavelength ranging from 8 nm to 20 nm are evident in STEM images of the Ca-rich dolomite samples (Figure 1). The observed features are similar to the lamellae in a Ca-rich dolomite first noted by Reeder et al. (1978). The lamellae are shown to arise from fluctuations in calcium content, since they are visible in both BF and HAADF images (Figure 1). In the diffraction pattern (Figure 2), extra reflections occur midway between the 000 and the $(1\bar{1}2)^*$ spot and between 000 and $(110)^*$. Note that extra reflections are streaked and parallel to $(110)^*$. According to the Fast Fourier Transform (FFT) patterns of the bright field image (Figure 3A), these extra reflections correspond to areas with lamellae rather than areas free of lamellae. However, these "*c*" reflections do not exist in the FFT patterns of either areas in the HAADF image (or Z-contrast image) that uses high-angle scattered incoherent electrons (Figure 3B).

Chemical lamellae were also observed in a high-resolution TEM image (Figure 4A) of the dolomite crystal viewed along the [010] zone axis (Figure 4B). The orientations of these lamellae vary from (001) to ~ $(\bar{1}01)$. Again, extra "*c*" reflections were found associated with some (but not all) of these lamellae that were not observed in lamellae-free regions (Figure 4C). The additional reflections were observed at $\frac{1}{2}(1\bar{1}02)^*$ and $\frac{1}{2}(104)^*$. These extra reflections are those reported as "*c*" reflections in previous TEM works of rhombohedral carbonates (Reeder and Wenk, 1979; Reeder, 1981; Van Tendeloo et al., 1985; Reksten, 1990a; Wenk et al., 1991; Fouke and Reeder, 1992; Schubel et al., 2000). The fact that "*c*" reflections were not always observed for lamellae implies that the additional Ca was not always ordered.

It was proposed by Larsson and Christy (2008) that the "*c*" reflections in the diffraction pattern can be generated by superposition of diffraction from the host dolomite crystal and that from inclusions of material with similar cell parameters but the disordered calcite structure, in an orientation related to the host by (104) twinning (Figure 5A). Similarly, we can produce the "$c$" reflections along the (104)* and ($\bar{1}$02)* directions in the diffraction pattern by superimposing the diffraction from the host dolomite and that from calcite lamellae that have ($\bar{1}$02) twin-like relationship with the host dolomite (Figure 5B). The unit cell parameters of lamellae are the same as those of the host dolomite.

In Z-contrast image of the Ca-rich dolomite crystal from the same sample as shown in the HRTEM image, similar yet more regular bright linear features or modulations have been observed (Figure 6A). The orientations of the lamellae (or chemical modulations) are parallel to (101) ~ (104), mostly in the range between (205) and (104), with a few exceptions, such as parallel to ($\bar{1}$08). The distance between successive lamellae was 7-30 nm. No extra spots such as "$c$" reflections were observed in FFT patterns from Z-contrast images of the lamellar regions, which suggests that the additional $Ca^{2+}$ ions do not form a superstructure when projected down the viewing direction. Intensity line profiles were taken parallel to the traces of ($\bar{1}$02) planes in order to estimate the variation in Ca:Mg ratio across the bright lamellae, making use of the $Z^{1.7}$ dependence of intensity in the Z-contrast image. Viewing along [010] direction, Ca and Mg cation layers alternate along the ($\bar{1}$02) traces in ideal dolomite. According to Figure 6B, the intensity associated with the Mg layers increases inside the linear features, which indicates the replacement of $Mg^{2+}$ by $Ca^{2+}$. Assuming that the dolomite outside the lamellae is nearly stoichiometric, we estimated the composition of the lamellae qualitatively. The host dolomite has 0~3% excess of $CaCO_3$, and the lamellae have compositions ranging from $Ca_{0.85}Mg_{0.15}CO_3$ to $Ca_{0.70}Mg_{0.30}CO_3$. FFT patterns from the lamellae in some areas show weak $b$ reflections such as (003). These weak spots are not from disordered calcite with $R\bar{3}m$ symmetry, but from areas overlapped with the host dolomite. The lamellae overlapped with were not used for line profile analysis of the Mg-calcite lamellae. All the compositions of the Mg-calcite compositions are based on the line profiles from the lamellae with $R\bar{3}c$ symmetry only.

**DISCUSSION**

The "*c*" type reflections usually accompanying the modulated microstructures are very common in Ca-rich dolomite. In this study, we found two of the three different kinds of "*c*" reflections, and as in previous work (Fig. 7c of Reeder 1981), we found that "*c*" reflections along the (104)* and ($\bar{1}$02)* directions or along (110)* and (1$\bar{1}$2)* directions exist in the same diffraction pattern. It has been reported that the "*c*" reflections can be either commensurate or incommensurate with the host structure (Schubel et al., 2000). It has been further proposed that "*c*" reflections form in domains with an ideal composition of $Ca_{0.75}Mg_{0.25}CO_3$ due to ordering of Mg and excess Ca in (001) planes, which doubles the periodicity in the *a* direction (Van Tendeloo et al., 1985). In the alternative model of Larsson and Christy (2008), "*c*" reflections arise not from additional cation ordering, but due to superposition of diffraction patterns from a dolomite host and nanoscale calcite inclusions that are oriented in a (104) twin relationship to the host. We have shown that a similar model with ($\bar{1}$02) as the twin plane can produce some of the "*c*" reflections seen in this study. In the model of Larsson and Christy (2008), multiple scattering by matrix and twinned nanodomains completes the extra weak "*c*" reflections. Although "*c*" reflections appear in SAED patterns and FFT patterns of the HRTEM image and bright-field STEM images, they do not appear in FFT patterns of HAADF images (Z-contrast images). Note that Z-contrast imaging uses high-angle scattered and incoherent electrons and therefore avoids multiple diffraction problems from the overlapped twinned crystals. Conversely, electron diffraction and bright-field imaging (HRTEM and STEM BF imaging) uses low-angle scattered and coherent electrons that will result in multiple diffraction from any overlapped twin lamellae that are present. We deduce from the difference in FFT patterns for the different image types that observed "*c*" reflections arise from nanodomains of magnesian calcite in a twinned orientation relative to a dolomite host, and that have anomalous cell parameters similar to those of the dolomite host, as in Larsson and Christy (2008). If unit cell parameters of the magnesian calcite lamellae (especially coarse lamellae) are larger than those of the dolomite host, positions of the "*c*" reflections will be off the center, which were observed in a Ca-rich dolomite from the Latemar buildup (Schubel, et al., 2000).

Modulation in some dolomites has been interpreted as due to strain associated with high-Ca domains that formed by exsolution or during growth (Fouke and Reeder, 1992; Reeder, 1992). However, our STEM study shows that contrast results purely from the composition difference between Ca-rich lamellae and dolomite matrix. We did not find any evidence of growth zoning, and interpret the modulation to have arisen from exsolution during diagenesis. Initially, extra Ca ions substitute for Mg on the Mg layers of the dolomite structure at low temperature (Fig. 7A). These Ca ions then migrate to form lamellae that are oriented parallel to planes such as (110) or (104), probably to minimize interfacial strain with the host dolomite (Figure 7B, 7C). Carbonate ions may also re-orient so as to put these lamellae in a twinned orientation and adjust the local cell parameters to fit the host dolomite lattice. The lamellae are metastable, however, and given time or exposure to higher temperature, further cation migration and carbonate re-orientation will occur so that coarse exsolution lamellae of calcite are formed parallel to (001) of dolomite (Fig.

7D). Initial compositional difference in different sectors (Reeder and Prosky, 1986; Fouke and Reeder, 1992) may also affect orientation difference of the exsolution lamellae.

## ACKNOWLDEGEMENTS

This work is supported by NASA Astrobiology Institute (N07-5489), NSF (EAR-095800), and U.S. Department of Energy (DE-FG02-09ER16050). We wanted to thank Dr. John Fournelle for providing the dolomite standard.

**Figures and Captions**

**Figure 1**

Bright field (A) and dark field (B) STEM images of the calcite lamellae in the same area of the Ca-rich dolomite. Strain contrast is evident in the BF image, but not in the DF image (B) due to collecting coherent electrons and incoherent electrons using different detectors.

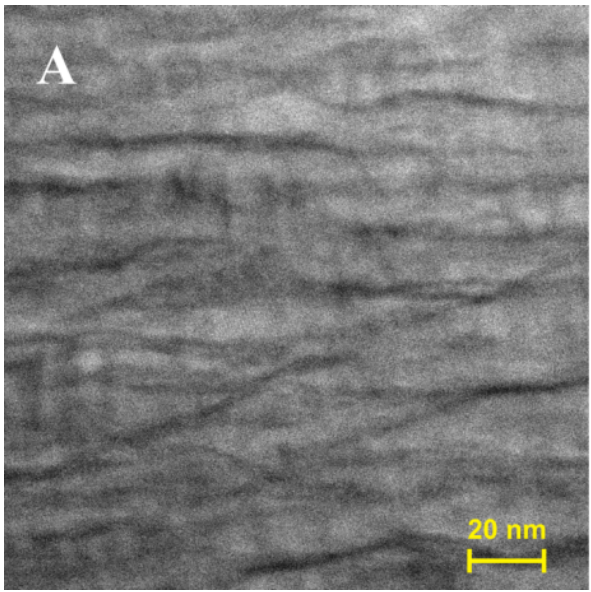 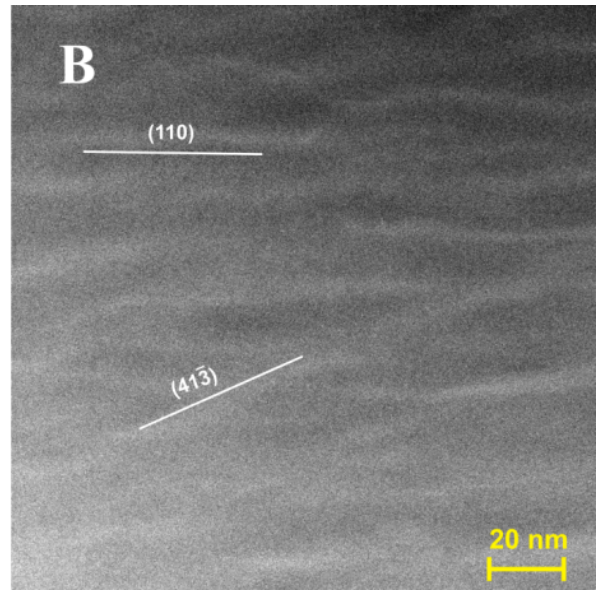

**Figure 2**

**Diffraction pattern of [$\bar{1}$11] zone axis shows the c reflections along the (1$\bar{1}$2)* and (110)* directions in the Ca-rich dolomite.**

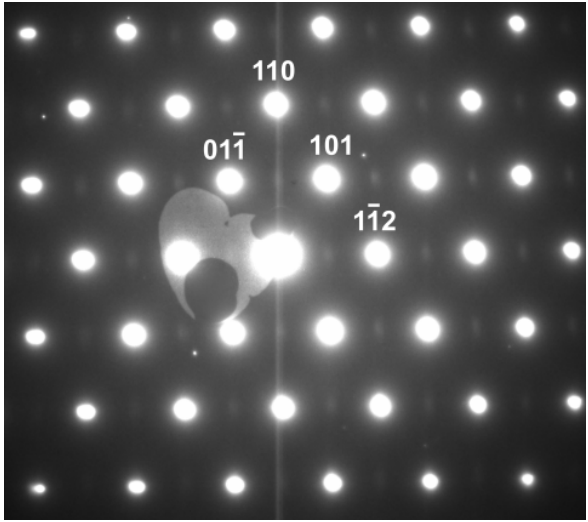

**Figure 3.**

A (Top): The extra "c" reflections along (1$\bar{1}$2)* and (110)* directions exist in the FFT pattern (1) from the lamellar domains in the bright field image, but not in FFT pattern (2) of the host dolomite.

B (Bottom): No "c" reflections in the FFT pattern (1) from the lamellar domains in Z-contrast image.

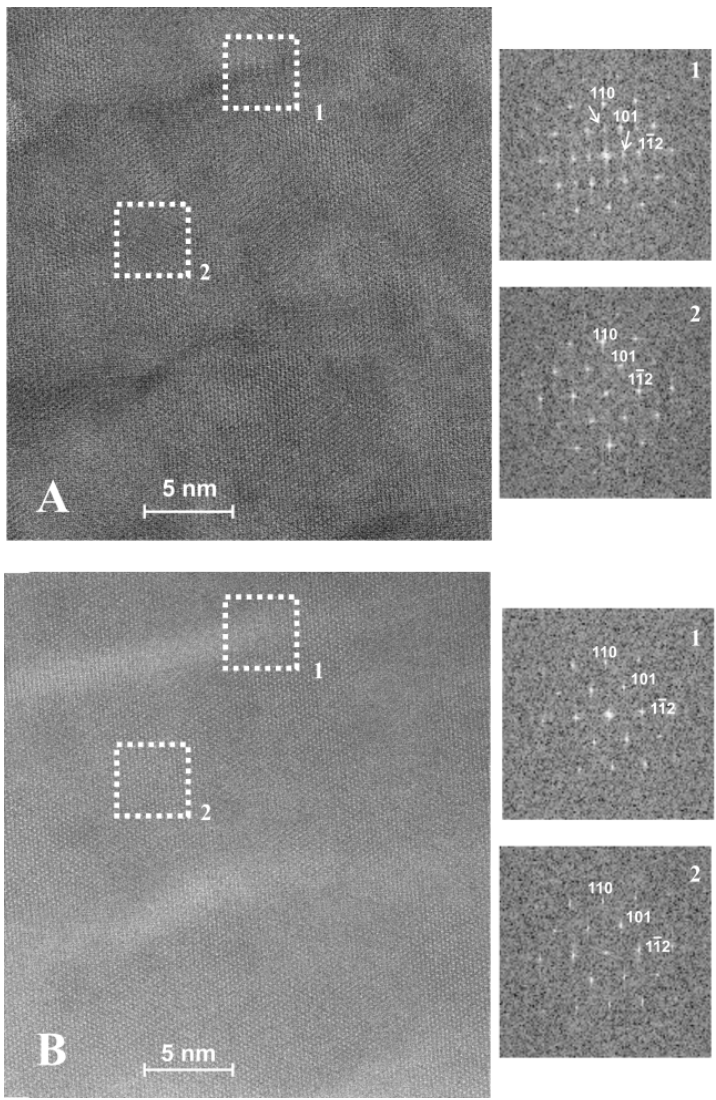

Figure 4

A: A [010]-zone axis HRTEM image of the Ca-rich dolomite shows the calcite exsolution lamellae. B. FFT pattern of the image showing "*c*" reflections along (104)* and (102)* (indicated by arrows). C: Extra "*c*" reflections exist in the FFT patterns from areas with overlapping features (e.g. region 1). The FFT patterns from lamellae with sharp boundaries do not show "*c*" reflections (region 2 and 3).

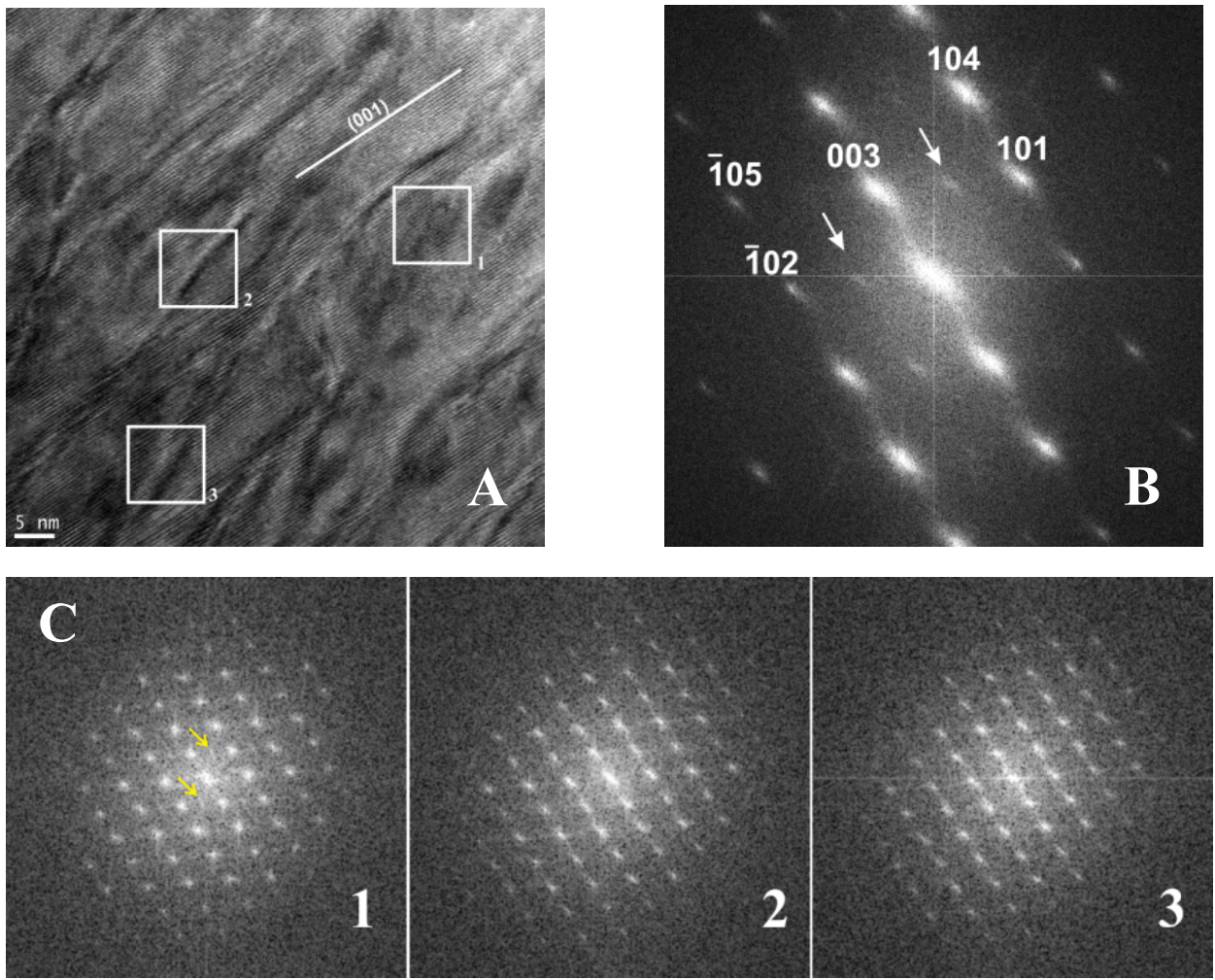

Figure 5

A: A "twin" model producing the "c" reflections along ($\bar{1}$02)* and (104)*directions in the diffraction pattern by superposition of the diffraction of the dolomite host and that of the calcite (104) "twin" (modified from Larsson and Christy, 2008). The calcite "twin" has the same unit cell parameters as the dolomite host. The "m" plane represents the (104) twin plane. B: The overlap of the diffraction pattern of the dolomite host and that of the calcite ($\bar{1}$02) twin may also result in the "c" reflections along ($\bar{1}$02)* and (104)*directions.

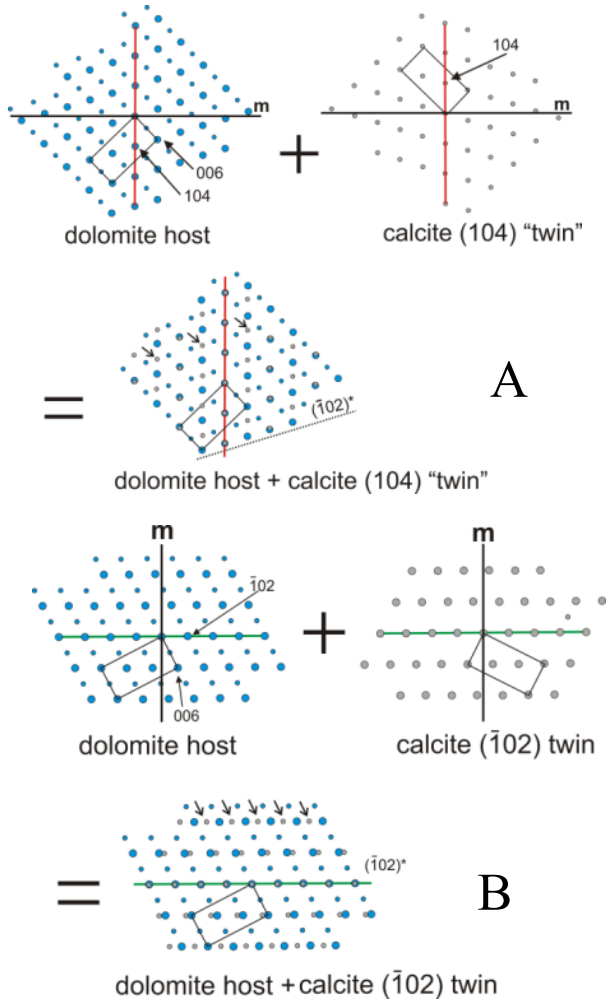

**Figure 6**

A (Top): Two line traverses '1' and '2' have been taken parallel to ($\bar{1}02$) trace in order to examine composition variation at atomic resolution. Compare FFT patterns from a lamella (a), lamella overlapped with dolomite host (b), and the dolomite host (c). (003) reflection (arrowed) in FFT pattern (b) is from the dolomite host. B: (bottom): An intensity profile of line '1' as shown in (A).

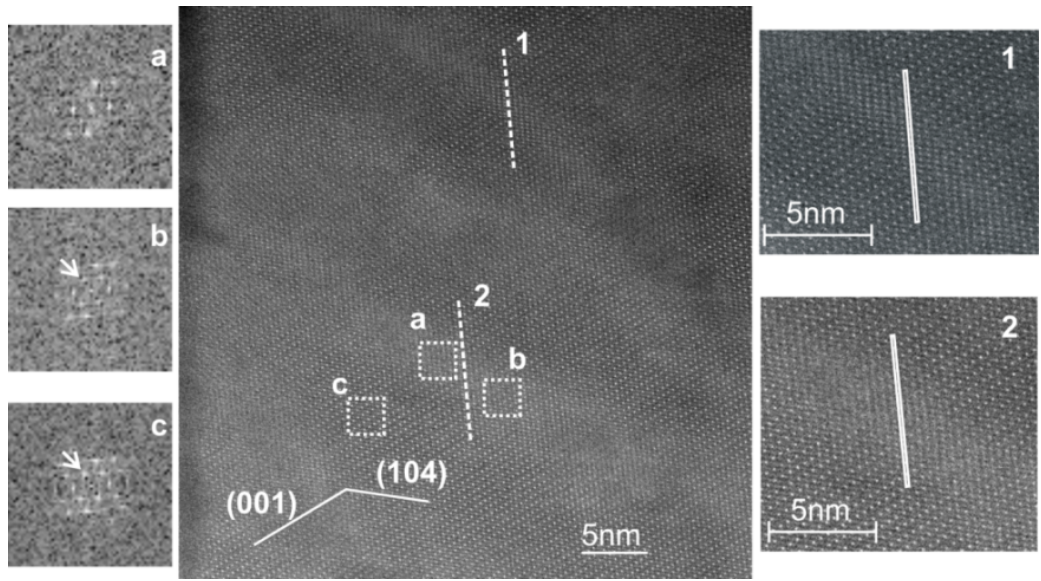

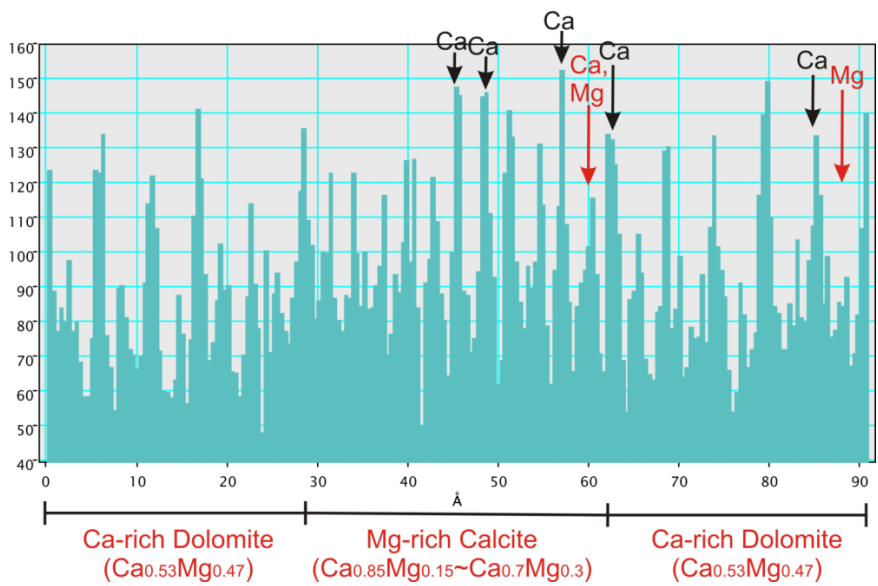

Ca-rich Dolomite ($Ca_{0.53}Mg_{0.47}$)  Mg-rich Calcite ($Ca_{0.85}Mg_{0.15} \sim Ca_{0.7}Mg_{0.3}$)  Ca-rich Dolomite ($Ca_{0.53}Mg_{0.47}$)

Figure 7

Proposed models for the formation of Ca-rich exsolution lamellae. (A) Initially, extra Ca²⁺ are incorporated into the $Mg^{2+}$-layers in dolomite structure. (B) These extra $Ca^{2+}$ migrate within the $Mg^{2+}$-layers and concentrate in linear regions forming exsolution lamellae parallel to (110). (C) Exsolution lamellae parallel to (104). (D) the exsolution lamellae parallel to basal plane (001). Exsolution lamellae in (104) and to ($\bar{1}$02) twin-like relationship with the dolomite host are also schematically proposed in E and F respectively.

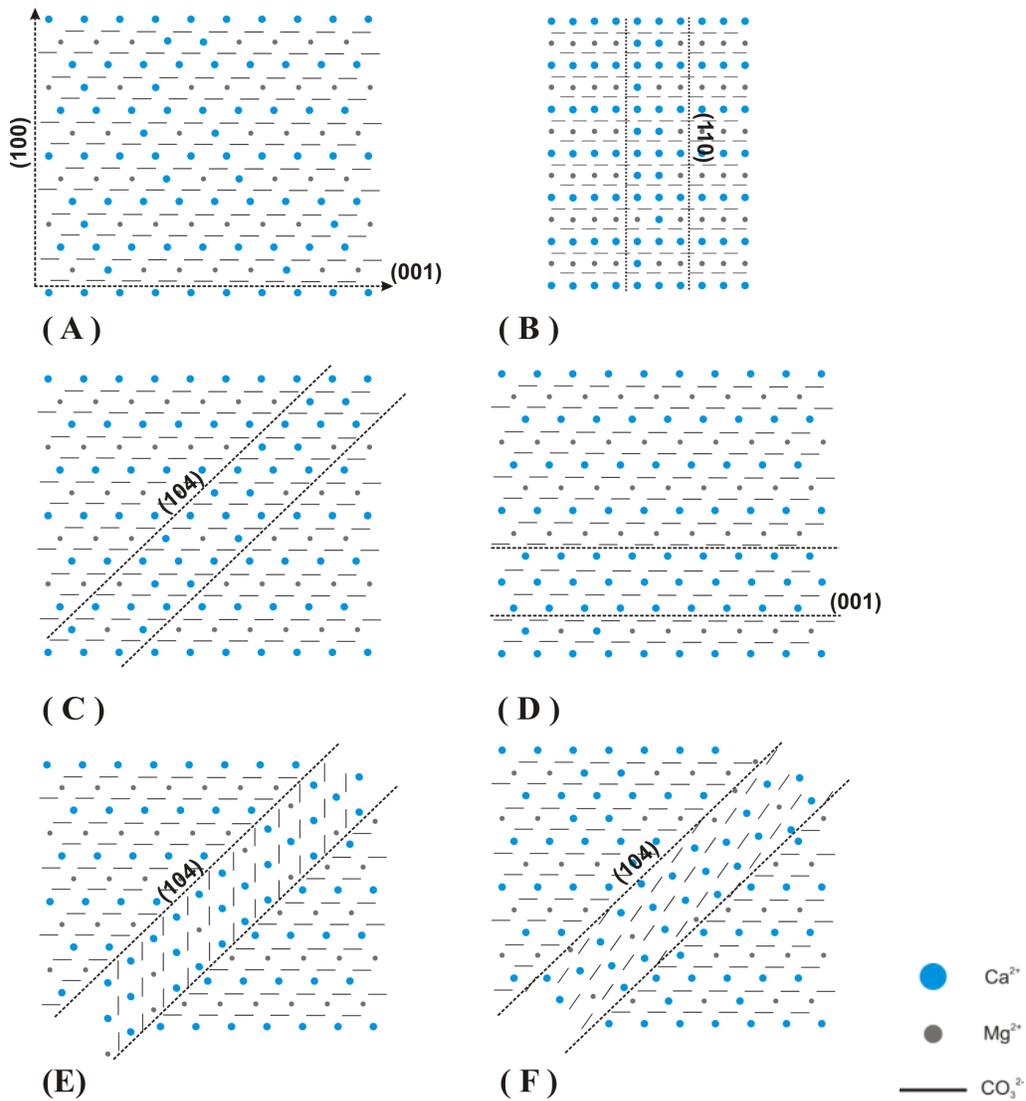